\documentclass[a4paper,11pt]{article}
\pdfoutput=1 

\usepackage{jcappub} 
                     
\usepackage[T1]{fontenc} 

\newcommand{\lcdm}{$\Lambda$CDM}
\def\gsim{\mathrel{\raise.3ex\hbox{$>$\kern-.75em\lower1ex\hbox{$\sim$}}}}
\def\lsim{\mathrel{\raise.3ex\hbox{$<$\kern-.75em\lower1ex\hbox{$\sim$}}}}

\title{\boldmath Gravitational potential wells and the cosmic bulk flow}

\author[a]{Abhinav Kumar,}
\author[a]{Yuyu Wang,}
\author[a]{Hume A. Feldman}
\author[b]{and Richard Watkins}

\affiliation[a]{Department of Physics \& Astronomy, University of Kansas, Lawrence, KS 66045, USA.}
\affiliation[b]{Department of Physics, Willamette University, Salem, OR 97301, USA.}

\emailAdd{abhispace@gmail.com}
\emailAdd{yuyuwang@ku.edu}
\emailAdd{feldman@ku.edu}
\emailAdd{rwatkins@willamette.edu}

\abstract{The bulk flow is a volume average of the peculiar velocities and a useful probe of the mass distribution on large scales. The gravitational instability model views the bulk flow as a potential flow that obeys a Maxwellian Distribution. We use two N-body simulations, the LasDamas Carmen and the Horizon Run, to calculate the bulk flows of various sized volumes in the simulation boxes. Once we have the bulk flow velocities as a function of scale, we investigate the mass and gravitational potential distribution around the volume. We found that matter densities can be asymmetrical and difficult to detect in real surveys, however, the gravitational potential and its gradient may provide better tools to investigate the underlying matter distribution. This study shows that bulk flows are indeed potential flows and thus provides information on the flow sources. We also show that bulk flow magnitudes follow a Maxwellian distribution on scales $>10\ h^{-1}$ Mpc.}


\begin{document}

\maketitle
\flushbottom

\section{Introduction}

The standard framework for the formation of the large-scale structure of the Universe is assumed to be governed by the gravitational instability model \citep{BarBonKai1986,MukFelBra1992,EisHu1998}; observational signatures of the model have been reported \citep{bispec1,bispec2,Verde2002} that confirm this scenario. The remarkable success of this model to predict and explain the formation of cosmic structure paved the way to the development of the Cosmological Constant Cold Dark Matter (\lcdm) scenario that agrees extremely well with a whole slew of observations, such as the Cosmic Microwave Background (CMB) fluctuations \citep{wmap9,Planckparameters2013}, and redshift surveys \citep{BlaBriCsa2003,EisZehHog2005,ColPerPea2005,JonReaSau2009,ThoAbdLah2011,SDSS811,SDSSIIIBAO2012}. Many of these observations, however, probe either high redshift tracers or light distribution ones which may be biased tracers of the current mass distribution. There is one probe, however, that dynamically traces the mass distribution, especially at small redshifts, and that is the cosmic velocity field, and more particularly in this paper, the bulk flow (e.g.  \citep{DreFab1990,CouFabDre1993,FelWat1994,BahGraCen1994,Hudson1994,Willick1994,WatFel1995,BahOh1996,Watkins1997,GioHaySal1998,HudSmiLuc1999,Willick1999,CouWilStr2000} and many others).

The bulk flow (BF) is the average of peculiar velocities in a volume \citep{WatFel2007,DavScr2014,Nusser2014}. Peculiar velocities  specify the velocities of galaxies towards mass concentrations and thus are good tracers of matter and the corresponding gravitational potentials \citep{Kashlinsky1991,BosJiaHea2014}. They provide us an important tool to discern the reason for cosmic flows of matter since gravitational potentials on these scales are dominated by dark matter. Citation \citep{Komatsu2009} has shown that these peculiar velocities agree with the $\Lambda$CDM model of the universe. Other studies have shown similar results as well \citep{Pike2005,Abate2009,NusBraDav2011,NusDav2011}.

However, some estimations of BF show some tension with the \lcdm\ model. Recent studies (e.g. \citep{WatFelHud2009,FelWatHud2010,MaGorFel2011,Turnbull2012,MacFelFer2012}) are inconsistent with \lcdm\ at $2.5- 3\sigma$ confidence level. If these estimations are not due to large scale flows, then the reason for the discrepancies may come from the uncertainty in original data, distance estimation methods and systemic uncertainties. Some of those uncertainties may be hard to control \citep{WatFel2014,WatFel2015}.

A popular method to estimate the BF is the Maximum Likelihood Estimate (MLE) \citep{Kaiser1988}. However, the MLE method has two major shortcomings \citep{WatFelHud2009,FelWatHud2010,MacFelFer2011,WatFel2015}: first, each survey probes the power spectrum in a unique way which makes it very difficult to directly compare different surveys; second, the MLE window function for a given survey is fixed, which means that the effective scale from which the results are estimated is particular to the survey and cannot be changed without discarding some of the data. Further, \citep{Nusser2014} also shows the inconsistency of the MLE. Recently, a more consistent and, in fact, optimal estimation method was introduced to address these problems, called the Minimum Variance (MV) method \citep{WatFelHud2009,FelWatHud2010}, and was tested further by citation \citep{AgaFelWat2012} (for a recent implementation of the MV method see \citep{HonSprStaScr2014,ScrDavBlaSta2015}).

Various studies have indicated that the BF magnitudes fit Maxwellian distribution (e.g. \citep{Li2012}), which is consistent with the expectation that  structure formation is the outgrowth of gaussian initial conditions via gravitational instability. Some possible sources of the BF are reviewed by citation \citep{RatKovItz2013}; they include attractors \citep{Dressler1987,Lynden1989}, super-horizon tilt \citep{Turner1992,KasAtrKoc2008,MaGorFel2011}, over-dense regions resulting from bubble collisions \citep{Larjo2010} or induced by cosmic defects \citep{Fialkov2010}. In recent studies, some new possible sources have been proposed: a combination of over- and under-dense region \citep{TulShaKar2008,FelWatHud2010} and void asymmetries in the cosmic web \citep{Bland2014}.

In this paper we use N-body simulations to revisit and test the basic assumptions made to analyze and understand survey data for cosmic flows. More specifically, the simulation data will provide us with both the velocity distributions and the sources of BF. This will enable us to determine whether the assumptions made regarding large-scale cosmic flows should be modified.

The paper is organized as follows: In section~\ref{sec:nbody} we describe the numerical simulations we use. In section~\ref{sec:weight} we discuss the method we use to estimate the BF. In section~\ref{sec:bulkflow} we show the BF estimation results and statistics. In section~\ref{sec:potential} we discuss the possible sources of BF. The results and conclusions are presented in section~\ref{sec:conclusions}.

\section{N-body simulations}
\label{sec:nbody}

In the following we use two independent suites of N-body simulations. The first are the Carmen Boxes of Large Suite of Dark Matter Simulations (LD-Carmen) \citep{LasDamas}, that contain positions and velocities of galaxies with equal mass.\footnote{http://lss.phy.vanderbilt.edu/lasdamas/} The other is the Horizon Run (HR) \citep{HR} simulation which uses Zel'dovich approximation \citep{ZA70,ZA} to generate initial conditions, containing positions, velocities and masses of galaxies, and is designed to model the SDSS observations. The parameters for LD-Carmen and HR simulations are shown in Table~\ref{LDHR}.

The LD-Carmen simulation uses a parallel friends-of-friends (FOF) code \citep{DavEfsFre1985} to identify bound groups of dark matter particles (halos) with the Ntropy framework \citep{GarConMcB2007}. The initial conditions are generated by the 2LPT code \citep{CroPueSco2006}, which employs second-order Lagrangian perturbation theory. The simulations include 41 Gpc$^3$ boxes with approximately 1.3 million mock galaxies in each. Following the Gaussian-weighted procedure we describe in section~\ref{sec:weight}, we extract $1,000$ mock surveys from each of the 41 LD-Carmen boxes to estimate a total of $41,000$ BF's. We also extract $5,000$ bulk flow catalogues from the HR simulation. The simulations we use differ from each other by small deviations of the initial conditions and the size and density of the data. It is also important to note that particle mass and softening also differ in the two simulations, to some extent owing to different initial and final redshifts (see Table~\ref{LDHR}).

\begin{table}
\caption{ The cosmological parameters and information of LD-Carmen and HR simulation}
\centering
{
\begin{tabular}{lcc}
&&\\
Parameters & LD-Carmen & HR\\
\hline
Matter density, $\Omega_m$ & 0.25 & 0.26\\
Cosmological constant density, $\Omega_\Lambda$ & 0.75 & 0.74\\
Baryon density, $\Omega_b$ & 0.04 & 0.044\\
Hubble parameter, $h$ ($100 km s^{-1} Mpc^{-1}$) & 0.7 & 0.72\\
Amplitude of matter density fluctuations, $\sigma_8$ & 0.8 & 0.794\\
Primordial scalar spectral index, $n_s$ & 1.0 & 0.96\\
Box size ($h^{-1}Mpc$) & 1000 & 6592\\
Number of particles & $1120^3$ &$4120^3$\\
Initial redshift, $z_0$ & 49 & 23\\
Redshift, $z$ & 0.13 & 0\\
Particle mass, $m_p$ ($10^{10} h^{-1} M_{\odot}$) & 4.938 & 29.6\\
Softening, $f_c$ ($h^{-1}kpc$) & 53 & 160\\
\end{tabular}%
}
\label{LDHR}
\end{table}

\section{The Gaussian-weighted bulk flow}
\label{sec:weight}

We calculate the Gaussian and tophat BF's using the full three dimensional velocity vectors directly.

We define the $i^{\rm th}$ component of the BF ($V_i$) to be
\begin{equation}
V_i =\frac{\sum\limits_{n=1}^N S_{i,n}f_n(r)}{\sum\limits_{n=1}^N f_n(r)}\, ,
\label{eq:bf}
\end{equation}
where $S_{i,n}$ is the $i^{th}$ peculiar velocity component of the $n$th galaxy, and $f(r)$ is a Gaussian radial distribution function
\begin{equation}
f(r) = e^{-r^2/2R_G^2}\, ,
\end{equation}
where $r$ is the distance from galaxy to the center of Gaussian ball and $R_G$ is a measure of the depth of the survey.

Applying this method to the simulation data, for each LD-Carmen box we pick 1,000 random points and estimate the Gaussian-weighted BF for each.
As a consistency check, we  compare it with the one suggested by citation \citep{Nusser2014}:
\begin{equation}\label{eq:Nusser}
\textbf{B}(r)=\frac{3}{4\pi r^{3}} \int_{0}^{r} \textbf{v}(\textbf{r}') d^{3}r'\ ,
\end{equation}
where $\textbf{B}(r)$ is the BF vector, $r$ is the radius of the tophat sphere which is twice the size of the $R_G$ in Gaussian-weighted method (see Eq.~\ref{eq:bf}), and $\textbf{v}(\textbf{r}')$ is the peculiar velocity vector. The BF's calculated with Eq.~\ref{eq:bf} and the ones in Eq.~\ref{eq:Nusser} give virtually identical results as long as the radii are corrected for the difference in the selection functions. 

\section{Bulk flow distribution}
\label{sec:bulkflow}

\begin{figure}
\centering
\includegraphics[scale=1]{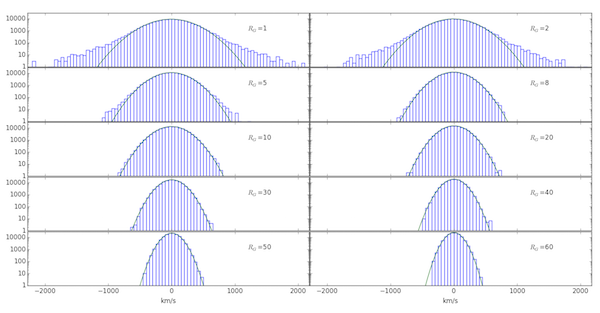}
\caption{\label{fig:velocity} The Logarithmic distribution of the BF components (41,000 LD-Carmen mock surveys, three components each) with $R_G=1-60$ $h^{-1}$Mpc radius. Bin widths equal to 50 km/s for all the sub-graphs. We can see clearly that the distribution is Gaussian (Solid line) for $R_G> 8$ $h^{-1}$Mpc.}
\end{figure}

Since $|V|^2=\sum\limits_i V_i^2$, and each component of the BF ($V_i$) has a Gaussian distribution, the BF magnitude should have a Maxwell-Boltzmann distribution (Maxwellian) instead of normal distribution. Figure~\ref{fig:velocity} shows the distribution of BF components. The figure shows clearly that for large $R_G$ the distribution is Gaussian, but for smaller radii, it is not. That is exactly what we expect given the small-scale nonlinearities in the velocity distribution.

To verify the BF magnitude distribution is Maxwellian distribution, we calculate its probability density function by using the least square method.

The Maxwellian distribution function is
\begin{equation}\label{eq:MW}
P(V)=\sqrt{\frac{2}{\pi}}\frac{V^2}{a^3}e^{-\frac{V^2}{2a^2}}\ ,
\end{equation}
where $a$ is the distribution parameter, $a=\sqrt{kT/m}$, derived from Maxwell-Boltzmann distribution function:

\begin{equation}\label{MBW}
P(V)=\sqrt{\left(\frac{m}{2\pi kT}\right)^3} 4\pi V^2 e^{-\frac{mV^2}{2kT}}\ .
\end{equation}

\begin{figure}
  \centering
  \includegraphics[scale=1]{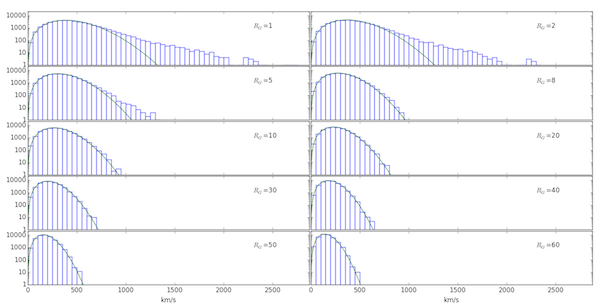}
  \caption{The Logarithmic distribution of the BF magnitude (41,000 LD-Carmen mock surveys) with varies radius. The distribution is Maxwellian (Solid line) for $R_G> 8h^{-1}$Mpc.
  \label{fig:BFmaxwell}
  }
\end{figure}

From Eq.~\ref{eq:MW} we get
\begin{equation}
\mu=2a\sqrt{\frac{2}{\pi}}\, ,\qquad \sigma=a^2\frac{(3\pi-8)}{\pi}\ ,
\end{equation}
where $\mu$ and $\sigma$ are the mean and variance of the distribution, respectively.
The mean ($\mu$) decreases as the survey depth ($R_G$) increases, that is, the BF of a volume decreases with increasing size, as expected.  This trend is also shown by figure~\ref{fig:BFmaxwell}. Here also we see that for small $R_G$, the distribution is not Maxwellian, again, as expected due to small-scale nonlinearities. We also tested the BF distribution from HR simulation, calculated $5,000$ BFs, which also follow a similar pattern.

\section{Potential}
\label{sec:potential}

Since there is no power on scales larger than the box size, and the simulation boxes are not tilted, that is, the average velocity of all galaxies in each direction vanishes, and the velocity field is curl free, the expectation is that BF's are potential flows due to mass distribution around the volumes.

According to citation \citep{LSS}, the  peculiar velocity is
\begin{equation}\label{PeculiarV1}
\textbf{v} = \frac{Haf}{4\pi}\partial _i\int d^3 x' \frac{\delta(x')}{|\textbf{x}'-\textbf{x}|}
\end{equation}
where $H$ is the Hubble constant, $a$ is scale factor and $f$ is a function of $\Omega$ (density parameter) $f(\Omega)=\Omega^{0.55}$ \citep{Linder2005}, $\partial_i$ is the partial derivative with respect to 3-dimensional coordinate ($i=x$, $y$, $z$).

Since the mock galaxies in the simulations have identical mass, we calculate the velocity in units of $\frac{Haf}{4\pi}$. Thus

\begin{equation}\label{PeculiarV2}
\textbf{v} = \partial _i\frac{1}{|\textbf{x}'-\textbf{x}|}\ .
\end{equation}
Since the peculiar velocity in the simulation is irrotational, we can express it as a gradient of a scaler.
\begin{equation}
\textbf{v} \varpropto -\nabla \phi
\end{equation}
Thus $\phi$ is proportional to $\frac{1}{|\textbf{x}'-\textbf{x}|}$ and since it has the same form as the gravitational potential, it fits our expectation. The potential equation can be simplified to
\begin{equation}
U_g = - \frac{1}{r'}
\end{equation}
The potential of a chosen point could be expressed as,
\begin{equation}\label{poeq}
U_g = -\sum\limits_{n=1}^N{\frac{1}{r'_n}}\, ,
\end{equation}
where $r'_n$ is the distance to the $n$th galaxy.

\begin{figure}
\centering
\includegraphics[scale=1]{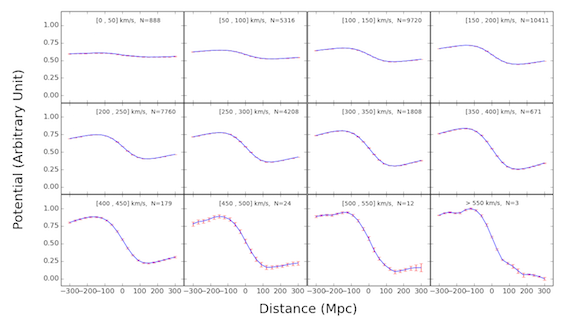}
\caption{\label{fig:LDpo}As the BF magnitude increases, potential slope about the center of the survey gets steeper. The potential has been rescaled to $[0,1]$. $N$ is the number of points in the simulations in the velocity range.}
\end{figure}

To make the analysis clearer, we rescaled $U_g$ to [0,1] by
\begin{equation}\label{poeqnorm}
U_{g_i} = \frac{U_{g_i} - \{U_g\}_{\rm min}}{\{U_g\}_{\rm max}-\{U_g\}_{\rm min}}\, .
\end{equation}
Where $U_{g_i}$ is potential of the $i^{\rm th}$ selected position and $\{U_g\}$ is the gravitational potential ensemble.

From Eqs.~(\ref{poeq}) and (\ref{poeqnorm}) we calculate the potential and its gradient along positive and negative BF direction centered about the Gaussian sphere center in a  [$-$300,300] $h^{-1}$Mpc range with step 30 $h^{-1}$Mpc.

\begin{figure}
\centering
\includegraphics[scale=1]{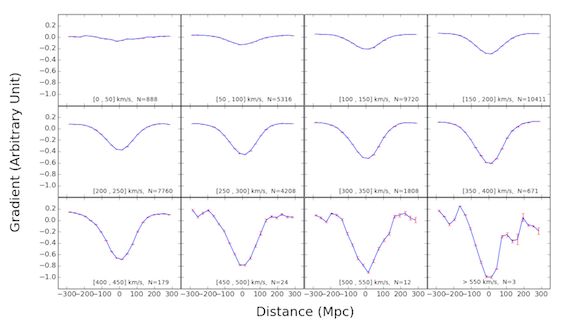}
\caption{\label{fig:LDgr}Same as figure~\ref{fig:LDpo}, but showing the gradient of the potential, rescaled to $-1$ minimum.}
\end{figure}

In figures~\ref{fig:LDpo} and~\ref{fig:LDgr} we show the average potential and its gradient for all BF velocities in different ranges of the BF magnitude, respectively. As can clearly be seen, there is a strong correlation between the BF's and the gradient of the potential, as expected. When we calculate the potential perpendicular to the BF direction, we get results consistent with no flows (as are the flows from the top left panels in figures~\ref{fig:LDpo} and~\ref{fig:LDgr}). 

The reason why the figures are symmetric about the central location of the volume is because these are averages of all flows in the magnitude range, the results suggest that generally we have an overdensity in the direction of the flow and an underdensity behind.

\begin{figure}
\centering
\includegraphics[scale=1]{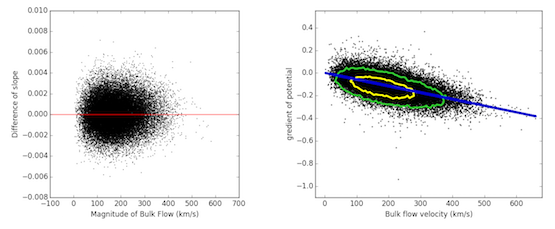}
\caption{\label{fig:Slopedist}Left panel: A scatter plot of the difference between the magnitude of the slope of the gradient in the direction of the bulk flow to the slope in the opposite direction. The red line notes the symmetric slopes. Right panel: We see a clear statistical correlation between BF magnitude and gradient of gravitational potential. As the BF magnitude increases, the gradient of the potential decreases. The contours contain 68\% and 95\% of the points and the line is the least square fit.}
\end{figure}

\begin{figure}
\centering
\includegraphics[scale=1]{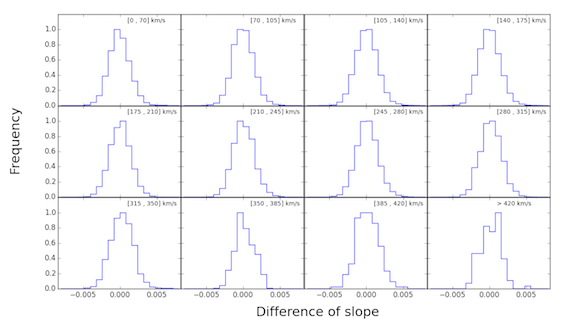}
\caption{\label{fig:Slopebin}The binned distribution of the gradient slope differences from figure~\ref{fig:Slopedist} left panel, rescaled to $[0,1]$. As can be expected, the distribution is roughly Gaussian.}
\end{figure}

However, for each individual volume, this is not necessarily the case. In figure~\ref{fig:Slopedist} left panel we show the difference between the magnitude of the slope of the potential gradient in the direction of the BF to that in the opposite direction. As can be clearly seen, each volume exhibits very different non-symmetric distribution. In figure~\ref{fig:Slopebin} we show the binned distribution of the gradient slope differences. As can be expected, the distribution is roughly Gaussian. This suggests that the reason for the flow is not a simple attractor but rather a complex superposition of over- and under-dense regions that combine to cause the flow. There is a clear statistical correlation between the BF magnitude and the gradient of the potential about the center of the volume for which we calculated the BF, as is shown in figure~\ref{fig:Slopedist} right panel. However, the picture is more complex when studying individual cases.

The BF magnitude from \citep{WatFelHud2009,FelWatHud2010,WatFel2015} increased with the radius ($R_G$) and was around 200 and 400 km/s at $R_G\approx20$ and $50\ h^{-1}$Mpc respectively. That is, the BF actually grew as a function of scale. We found above that the likelihood in \lcdm\ to have a 400 km/s BF magnitude on $50\ h^{-1}$Mpc scale to be about 2\% and as can be seen in Figures~\ref{fig:velocity} and~\ref{fig:BFmaxwell} a small percentage of the BF's are quite large. In the simulations we searched for  a rising BF magnitude as a function of scale as we found in the surveys. In other words, we tried to find a place in the simulation boxes that will emulate the same BF magnitude behavior on all scales ($10 \leq R_G \leq 60 h^{-1}$Mpc).

\begin{figure}
\centering
\includegraphics[scale=1]{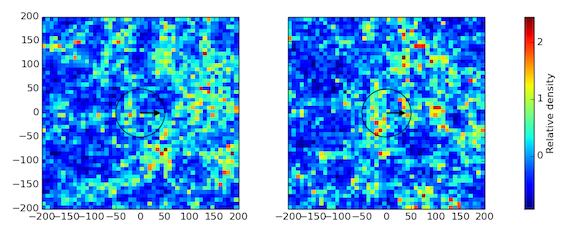}
\caption{\label{fig:DensityMap} Two representative slices of 20 and 50 $h^{-1}$Mpc BF planes of positions in the simulation that give similar BF behaviors to the one found in citation \citep{WatFel2014}. The black arrow and the black circle indicate the direction of the BF and radius of 50 $h^{-1}$Mpc scale, respectively.}
\end{figure}

We found two dozen of these type of regions among the 41k points we investigated. In Figs.~\ref{fig:DensityMap} we show two representative slices in the simulation box. The circle in each figure has a radius of 50 $h^{-1}$Mpc and the arrow shows the direction of the BF vector on this scale. Each slice was oriented so that the BF is to the positive x-direction. Each slice is 200$h^{-1}$Mpc thick.

As can be seen in the figures, the density is anisotropic on both small and large scales. On small scales the distributions are such that there is a small over-and under-dense regions close to the center of the volume which happen to be in the opposite direction of the large over- and under-dense regions on large scales. On all scales, the volume is attracted to the large over-dense regions in the positive x-direction and away from the large under-dense regions in the negative x-direction. On small scales the volume is also attracted to the close by over-dense regions in the negative x-direction and away from some under-dense regions close by in the positive x-direction which counters the large-scale flow and leads to slower overall flow on small scales.  

\begin{figure}
\centering
\includegraphics[scale=0.5]{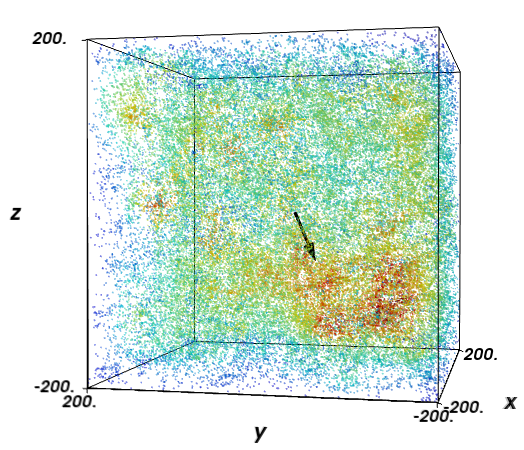}
\includegraphics[scale=0.6]{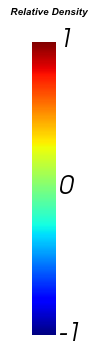}
\caption{\label{fig:DensityMap3} The galaxy distribution around the center of the volume analyzed in perspective. The shades represent the relative densities around each galaxy. }
\end{figure}

This suggests a very specific distribution of masses as well as large velocity shear that we may be able to detect in future surveys. Further, it shows that there is no one large source for the BF, but rather a complex distribution of high- and low-densities that gives rise to the flow. In figure~\ref{fig:DensityMap3} we show a perspective of the volume around the flow which shows similar behavior.  

\section{Conclusions}
\label{sec:conclusions}

The Bulk Flow statistic is a powerful probe of the mass distribution and the underlying gravitational potential on $>10\ h^{-1}$Mpc scales. We have investigated the BF's from numerical simulations and as expected, the distribution of BF components obey a Gaussian distribution, whereas the BF magnitude yields a Maxwellian distribution. The BF's magnitudes yield similar results to the ones suggested by citation \citep{Nusser2014}. As a check, we made sure that the BF distributions from LD-Carmen and HR simulations are similar and consistent with expectations.

We demonstrated that on scales $<10\ h^{-1}$Mpc, the velocity distribution is non Gaussian, whereas on larger scales it is. Further, we showed that the scalar gravitational potential field gives rise to the BF. Since our probe was the BF magnitudes, as long as we restrict our attention to large scales, we are sensitive mainly to linear flows and can approximate it as an irrotational potential flow. As such, we expect velocities to point towards mass concentration sources. The BF is expected to be a good tracer of the potential flow towards mass concentrations and away from underdense regions. 

In general, we may expect to locate high density regions in the direction of the BF, however, such a flow may also be a result of the relative differences between two opposite directions. Relatively high matter concentration in the direction of the flow and comparatively low density in the other direction may also produce such a result, as well as more complex distributions.
And thus, the observational search to find out the exact potential map or mass distribution may be very difficult to achieve. 

A large BF can have more than one over -and under-dense sources. Therefore, the gradient of gravitational potential or the BF may not point to over-dense region exactly. Complex geometrical distribution of mass may play a significant role here and prevent us from mapping the mass distribution with any precision, especially on large scales.

It is clear that although a BF of the magnitude ($\approx400$ km/s) found by some studies, are not likely ($<2$\% chance), they are not impossible either and can be driven by a mass distribution consistent with the $\Lambda$CDM scenario. To verify this observationally we must create accurate maps of the mass distribution on scales of 300-400 $h^{-1}$Mpc scales, a goal that may prove difficult to achieve in the near future.

\bibliographystyle{JHEP}
\bibliography{BulkFlow}

\providecommand{\href}[2]{#2}\begingroup\raggedright\begin{thebibliography}{10}

\bibitem{BarBonKai1986}
J.~M. {Bardeen}, J.~R. {Bond}, N.~{Kaiser}, and A.~S. {Szalay}, {\it {The
  statistics of peaks of Gaussian random fields}},  {\em \apj} {\bf 304} (May,
  1986) 15--61.

\bibitem{MukFelBra1992}
V.~F. {Mukhanov}, H.~A. {Feldman}, and R.~H. {Brandenberger}, {\it {Theory of
  cosmological perturbations}},  {\em \physrep} {\bf 215} (June, 1992)
  203--333.

\bibitem{EisHu1998}
D.~J. {Eisenstein} and W.~{Hu}, {\it {Baryonic Features in the Matter Transfer
  Function}},  {\em \apj} {\bf 496} (Mar., 1998) 605--614,
  [\href{http://arxiv.org/abs/astro-ph/9709112}{{\tt astro-ph/9709112}}].

\bibitem{bispec1}
R.~{Scoccimarro}, H.~A. {Feldman}, J.~N. {Fry}, and J.~A. {Frieman}, {\it {The
  Bispectrum of IRAS Redshift Catalogs}},  {\em \apj} {\bf 546} (Jan., 2001)
  652--664, [\href{http://arxiv.org/abs/astro-ph/0004087}{{\tt
  astro-ph/0004087}}].

\bibitem{bispec2}
H.~A. {Feldman}, J.~A. {Frieman}, J.~N. {Fry}, and R.~{Scoccimarro}, {\it
  {Constraints on Galaxy Bias, Matter Density, and Primordial Non-Gaussianity
  from the PSCz Galaxy Redshift Survey}},  {\em Physical Review Letters} {\bf
  86} (Feb., 2001) 1434, [\href{http://arxiv.org/abs/astro-ph/0010205}{{\tt
  astro-ph/0010205}}].

\bibitem{Verde2002}
L.~{Verde}, A.~F. {Heavens}, W.~J. {Percival}, S.~{Matarrese}, C.~M. {Baugh},
  J.~{Bland-Hawthorn}, T.~{Bridges}, R.~{Cannon}, S.~{Cole}, M.~{Colless},
  C.~{Collins}, W.~{Couch}, G.~{Dalton}, R.~{De Propris}, S.~P. {Driver},
  G.~{Efstathiou}, R.~S. {Ellis}, C.~S. {Frenk}, K.~{Glazebrook}, C.~{Jackson},
  O.~{Lahav}, I.~{Lewis}, S.~{Lumsden}, S.~{Maddox}, D.~{Madgwick},
  P.~{Norberg}, J.~A. {Peacock}, B.~A. {Peterson}, W.~{Sutherland}, and
  K.~{Taylor}, {\it {The 2dF Galaxy Redshift Survey: the bias of galaxies and
  the density of the Universe}},  {\em \mnras} {\bf 335} (Sept., 2002)
  432--440, [\href{http://arxiv.org/abs/astro-ph/0112161}{{\tt
  astro-ph/0112161}}].

\bibitem{wmap9}
C.~L. {Bennett}, D.~{Larson}, J.~L. {Weiland}, N.~{Jarosik}, G.~{Hinshaw},
  N.~{Odegard}, K.~M. {Smith}, R.~S. {Hill}, B.~{Gold}, M.~{Halpern},
  E.~{Komatsu}, M.~R. {Nolta}, L.~{Page}, D.~N. {Spergel}, E.~{Wollack},
  J.~{Dunkley}, A.~{Kogut}, M.~{Limon}, S.~S. {Meyer}, G.~S. {Tucker}, and
  E.~L. {Wright}, {\it {Nine-year Wilkinson Microwave Anisotropy Probe (WMAP)
  Observations: Final Maps and Results}},  {\em \apjs} {\bf 208} (Oct., 2013)
  20, [\href{http://arxiv.org/abs/1212.5225}{{\tt arXiv:1212.5225}}].

\bibitem{Planckparameters2013}
{Planck Collaboration}, P.~A.~R. {Ade}, N.~{Aghanim}, C.~{Armitage-Caplan},
  M.~{Arnaud}, M.~{Ashdown}, F.~{Atrio-Barandela}, J.~{Aumont},
  C.~{Baccigalupi}, A.~J. {Banday}, and et~al., {\it {Planck 2013 results. XVI.
  Cosmological parameters}},  {\em ArXiv e-prints} (Mar., 2013)
  [\href{http://arxiv.org/abs/1303.5076}{{\tt arXiv:1303.5076}}].

\bibitem{BlaBriCsa2003}
M.~R. {Blanton}, J.~{Brinkmann}, I.~{Csabai}, M.~{Doi}, D.~{Eisenstein},
  M.~{Fukugita}, J.~E. {Gunn}, D.~W. {Hogg}, and D.~J. {Schlegel}, {\it
  {Estimating Fixed-Frame Galaxy Magnitudes in the Sloan Digital Sky Survey}},
  {\em \aj} {\bf 125} (May, 2003) 2348--2360,
  [\href{http://arxiv.org/abs/astro-ph/0205243}{{\tt astro-ph/0205243}}].

\bibitem{EisZehHog2005}
D.~J. {Eisenstein}, I.~{Zehavi}, D.~W. {Hogg}, R.~{Scoccimarro}, M.~R.
  {Blanton}, R.~C. {Nichol}, R.~{Scranton}, H.-J. {Seo}, M.~{Tegmark},
  Z.~{Zheng}, S.~F. {Anderson}, J.~{Annis}, N.~{Bahcall}, J.~{Brinkmann},
  S.~{Burles}, F.~J. {Castander}, A.~{Connolly}, I.~{Csabai}, M.~{Doi},
  M.~{Fukugita}, J.~A. {Frieman}, K.~{Glazebrook}, J.~E. {Gunn}, J.~S.
  {Hendry}, G.~{Hennessy}, Z.~{Ivezi{\'c}}, S.~{Kent}, G.~R. {Knapp}, H.~{Lin},
  Y.-S. {Loh}, R.~H. {Lupton}, B.~{Margon}, T.~A. {McKay}, A.~{Meiksin}, J.~A.
  {Munn}, A.~{Pope}, M.~W. {Richmond}, D.~{Schlegel}, D.~P. {Schneider},
  K.~{Shimasaku}, C.~{Stoughton}, M.~A. {Strauss}, M.~{SubbaRao}, A.~S.
  {Szalay}, I.~{Szapudi}, D.~L. {Tucker}, B.~{Yanny}, and D.~G. {York}, {\it
  {Detection of the Baryon Acoustic Peak in the Large-Scale Correlation
  Function of SDSS Luminous Red Galaxies}},  {\em \apj} {\bf 633} (Nov., 2005)
  560--574, [\href{http://arxiv.org/abs/astro-ph/0501171}{{\tt
  astro-ph/0501171}}].

\bibitem{ColPerPea2005}
S.~{Cole}, W.~J. {Percival}, J.~A. {Peacock}, P.~{Norberg}, C.~M. {Baugh},
  C.~S. {Frenk}, I.~{Baldry}, J.~{Bland-Hawthorn}, T.~{Bridges}, R.~{Cannon},
  M.~{Colless}, C.~{Collins}, W.~{Couch}, N.~J.~G. {Cross}, G.~{Dalton}, V.~R.
  {Eke}, R.~{De Propris}, S.~P. {Driver}, G.~{Efstathiou}, R.~S. {Ellis},
  K.~{Glazebrook}, C.~{Jackson}, A.~{Jenkins}, O.~{Lahav}, I.~{Lewis},
  S.~{Lumsden}, S.~{Maddox}, D.~{Madgwick}, B.~A. {Peterson}, W.~{Sutherland},
  and K.~{Taylor}, {\it {The 2dF Galaxy Redshift Survey: power-spectrum
  analysis of the final data set and cosmological implications}},  {\em \mnras}
  {\bf 362} (Sept., 2005) 505--534,
  [\href{http://arxiv.org/abs/astro-ph/0501174}{{\tt astro-ph/0501174}}].

\bibitem{JonReaSau2009}
D.~H. {Jones}, M.~A. {Read}, W.~{Saunders}, M.~{Colless}, T.~{Jarrett}, Q.~A.
  {Parker}, A.~P. {Fairall}, T.~{Mauch}, E.~M. {Sadler}, F.~G. {Watson},
  D.~{Burton}, L.~A. {Campbell}, P.~{Cass}, S.~M. {Croom}, J.~{Dawe},
  K.~{Fiegert}, L.~{Frankcombe}, M.~{Hartley}, J.~{Huchra}, D.~{James},
  E.~{Kirby}, O.~{Lahav}, J.~{Lucey}, G.~A. {Mamon}, L.~{Moore}, B.~A.
  {Peterson}, S.~{Prior}, D.~{Proust}, K.~{Russell}, V.~{Safouris}, K.-I.
  {Wakamatsu}, E.~{Westra}, and M.~{Williams}, {\it {The 6dF Galaxy Survey:
  final redshift release (DR3) and southern large-scale structures}},  {\em
  \mnras} {\bf 399} (Oct., 2009) 683--698,
  [\href{http://arxiv.org/abs/0903.5451}{{\tt arXiv:0903.5451}}].

\bibitem{ThoAbdLah2011}
S.~A. {Thomas}, F.~B. {Abdalla}, and O.~{Lahav}, {\it {Excess Clustering on
  Large Scales in the MegaZ DR7 Photometric Redshift Survey}},  {\em Physical
  Review Letters} {\bf 106} (June, 2011) 241301--+,
  [\href{http://arxiv.org/abs/1012.2272}{{\tt arXiv:1012.2272}}].

\bibitem{SDSS811}
H.~{Aihara}, C.~{Allende Prieto}, D.~{An}, S.~F. {Anderson}, {\'E}.~{Aubourg},
  E.~{Balbinot}, T.~C. {Beers}, A.~A. {Berlind}, S.~J. {Bickerton},
  D.~{Bizyaev}, M.~R. {Blanton}, J.~J. {Bochanski}, A.~S. {Bolton}, J.~{Bovy},
  W.~N. {Brandt}, J.~{Brinkmann}, P.~J. {Brown}, J.~R. {Brownstein}, N.~G.
  {Busca}, H.~{Campbell}, M.~A. {Carr}, Y.~{Chen}, C.~{Chiappini},
  J.~{Comparat}, N.~{Connolly}, M.~{Cortes}, R.~A.~C. {Croft}, A.~J. {Cuesta},
  L.~N. {da Costa}, J.~R.~A. {Davenport}, K.~{Dawson}, S.~{Dhital}, A.~{Ealet},
  G.~L. {Ebelke}, E.~M. {Edmondson}, D.~J. {Eisenstein}, S.~{Escoffier},
  M.~{Esposito}, M.~L. {Evans}, X.~{Fan}, B.~{Femen{\'{\i}}a Castell{\'a}},
  A.~{Font-Ribera}, P.~M. {Frinchaboy}, J.~{Ge}, B.~A. {Gillespie},
  G.~{Gilmore}, J.~I. {Gonz{\'a}lez Hern{\'a}ndez}, J.~R. {Gott}, A.~{Gould},
  E.~K. {Grebel}, J.~E. {Gunn}, J.-C. {Hamilton}, P.~{Harding}, D.~W. {Harris},
  S.~L. {Hawley}, F.~R. {Hearty}, S.~{Ho}, D.~W. {Hogg}, J.~A. {Holtzman},
  K.~{Honscheid}, N.~{Inada}, I.~I. {Ivans}, L.~{Jiang}, J.~A. {Johnson},
  C.~{Jordan}, W.~P. {Jordan}, E.~A. {Kazin}, D.~{Kirkby}, M.~A. {Klaene},
  G.~R. {Knapp}, J.-P. {Kneib}, C.~S. {Kochanek}, L.~{Koesterke}, J.~A.
  {Kollmeier}, R.~G. {Kron}, H.~{Lampeitl}, D.~{Lang}, J.-M. {Le Goff}, Y.~S.
  {Lee}, Y.-T. {Lin}, D.~C. {Long}, C.~P. {Loomis}, S.~{Lucatello},
  B.~{Lundgren}, R.~H. {Lupton}, Z.~{Ma}, N.~{MacDonald}, S.~{Mahadevan},
  M.~A.~G. {Maia}, M.~{Makler}, E.~{Malanushenko}, V.~{Malanushenko},
  R.~{Mandelbaum}, C.~{Maraston}, D.~{Margala}, K.~L. {Masters}, C.~K.
  {McBride}, P.~M. {McGehee}, I.~D. {McGreer}, B.~{M{\'e}nard},
  J.~{Miralda-Escud{\'e}}, H.~L. {Morrison}, F.~{Mullally}, D.~{Muna}, J.~A.
  {Munn}, H.~{Murayama}, A.~D. {Myers}, T.~{Naugle}, A.~F. {Neto}, D.~C.
  {Nguyen}, R.~C. {Nichol}, R.~W. {O'Connell}, R.~L.~C. {Ogando}, M.~D.
  {Olmstead}, D.~J. {Oravetz}, N.~{Padmanabhan}, N.~{Palanque-Delabrouille},
  K.~{Pan}, P.~{Pandey}, I.~{P{\^a}ris}, W.~J. {Percival}, P.~{Petitjean},
  R.~{Pfaffenberger}, J.~{Pforr}, S.~{Phleps}, C.~{Pichon}, M.~M. {Pieri},
  F.~{Prada}, A.~M. {Price-Whelan}, M.~J. {Raddick}, B.~H.~F. {Ramos},
  C.~{Reyl{\'e}}, J.~{Rich}, G.~T. {Richards}, H.-W. {Rix}, A.~C. {Robin},
  H.~J. {Rocha-Pinto}, C.~M. {Rockosi}, N.~A. {Roe}, E.~{Rollinde}, A.~J.
  {Ross}, N.~P. {Ross}, B.~M. {Rossetto}, A.~G. {S{\'a}nchez}, C.~{Sayres},
  D.~J. {Schlegel}, K.~J. {Schlesinger}, S.~J. {Schmidt}, D.~P. {Schneider},
  E.~{Sheldon}, Y.~{Shu}, J.~{Simmerer}, A.~E. {Simmons}, T.~{Sivarani}, S.~A.
  {Snedden}, J.~S. {Sobeck}, M.~{Steinmetz}, M.~A. {Strauss}, A.~S. {Szalay},
  M.~{Tanaka}, A.~R. {Thakar}, D.~{Thomas}, J.~L. {Tinker}, B.~M. {Tofflemire},
  R.~{Tojeiro}, C.~A. {Tremonti}, J.~{Vandenberg}, M.~{Vargas Maga{\~n}a},
  L.~{Verde}, N.~P. {Vogt}, D.~A. {Wake}, J.~{Wang}, B.~A. {Weaver}, D.~H.
  {Weinberg}, M.~{White}, S.~D.~M. {White}, B.~{Yanny}, N.~{Yasuda},
  C.~{Yeche}, and I.~{Zehavi}, {\it {The Eighth Data Release of the Sloan
  Digital Sky Survey: First Data from SDSS-III}},  {\em \apjs} {\bf 193} (Apr.,
  2011) 29, [\href{http://arxiv.org/abs/1101.1559}{{\tt arXiv:1101.1559}}].

\bibitem{SDSSIIIBAO2012}
A.~J. {Ross}, W.~J. {Percival}, A.~G. {S{\'a}nchez}, L.~{Samushia}, S.~{Ho},
  E.~{Kazin}, M.~{Manera}, B.~{Reid}, M.~{White}, R.~{Tojeiro}, C.~K.
  {McBride}, X.~{Xu}, D.~A. {Wake}, M.~A. {Strauss}, F.~{Montesano}, M.~E.~C.
  {Swanson}, S.~{Bailey}, A.~S. {Bolton}, A.~M. {Dorta}, D.~J. {Eisenstein},
  H.~{Guo}, J.-C. {Hamilton}, R.~C. {Nichol}, N.~{Padmanabhan}, F.~{Prada},
  D.~J. {Schlegel}, M.~V. {Maga{\~n}a}, I.~{Zehavi}, M.~{Blanton},
  D.~{Bizyaev}, H.~{Brewington}, A.~J. {Cuesta}, E.~{Malanushenko},
  V.~{Malanushenko}, D.~{Oravetz}, J.~{Parejko}, K.~{Pan}, D.~P. {Schneider},
  A.~{Shelden}, A.~{Simmons}, S.~{Snedden}, and G.-b. {Zhao}, {\it {The
  clustering of galaxies in the SDSS-III Baryon Oscillation Spectroscopic
  Survey: analysis of potential systematics}},  {\em \mnras} {\bf 424} (July,
  2012) 564--590, [\href{http://arxiv.org/abs/1203.6499}{{\tt
  arXiv:1203.6499}}].

\bibitem{DreFab1990}
A.~{Dressler} and S.~M. {Faber}, {\it Confirmation of a large-scale,
  large-amplitude flow in the direction of the great attractor},  {\em \apj}
  {\bf 354} (May, 1990) 13--17.

\bibitem{CouFabDre1993}
S.~{Courteau}, S.~M. {Faber}, A.~{Dressler}, and J.~A. {Willick}, {\it
  {Streaming motions in the local universe - Evidence for large-scale,
  low-amplitude density fluctuations}},  {\em \apjl} {\bf 412} (Aug., 1993)
  L51--L54.

\bibitem{FelWat1994}
H.~A. {Feldman} and R.~{Watkins}, {\it {Theoretical expectations for bulk flows
  in large-scale surveys}},  {\em \apjl} {\bf 430} (July, 1994) L17--L20,
  [\href{http://arxiv.org/abs/astro-ph/9312038}{{\tt astro-ph/9312038}}].

\bibitem{BahGraCen1994}
N.~A. {Bahcall}, M.~{Gramann}, and R.~{Cen}, {\it {The motions of clusters and
  group of galaxies}},  {\em \apj} {\bf 436} (Nov., 1994) 23--32,
  [\href{http://arxiv.org/abs/astro-ph/9410061}{{\tt astro-ph/9410061}}].

\bibitem{Hudson1994}
M.~J. {Hudson}, {\it {Optical galaxies within 8000 km s$^{-1}$ - IV. The
  peculiar velocity field}},  {\em \mnras} {\bf 266} (Jan., 1994) 475--488.

\bibitem{Willick1994}
J.~A. {Willick}, {\it {Statistical bias in distance and peculiar velocity
  estimation. 1: The 'calibration' problem}},  {\em \apjs} {\bf 92} (May, 1994)
  1--31.

\bibitem{WatFel1995}
R.~{Watkins} and H.~A. {Feldman}, {\it {Interpreting New Data on Large-Scale
  Bulk Flows}},  {\em \apjl} {\bf 453} (Nov., 1995) L73+,
  [\href{http://arxiv.org/abs/astro-ph/9505038}{{\tt astro-ph/9505038}}].

\bibitem{BahOh1996}
N.~A. {Bahcall} and S.~P. {Oh}, {\it {The Peculiar Velocity Function of Galaxy
  Clusters}},  {\em \apjl} {\bf 462} (May, 1996) L49,
  [\href{http://arxiv.org/abs/astro-ph/9602149}{{\tt astro-ph/9602149}}].

\bibitem{Watkins1997}
R.~{Watkins}, {\it {The RMS peculiar velocity of clusters}},  {\em \mnras} {\bf
  292} (Dec., 1997) L59--L63,
  [\href{http://arxiv.org/abs/astro-ph/9710170}{{\tt astro-ph/9710170}}].

\bibitem{GioHaySal1998}
R.~{Giovanelli}, M.~P. {Haynes}, J.~J. {Salzer}, G.~{Wegner}, L.~N. {da Costa},
  and W.~{Freudling}, {\it {The Motions of Clusters of Galaxies and the Dipoles
  of the Peculiar Velocity Field}},  {\em \aj} {\bf 116} (Dec., 1998)
  2632--2643, [\href{http://arxiv.org/abs/astro-ph/9808158}{{\tt
  astro-ph/9808158}}].

\bibitem{HudSmiLuc1999}
M.~J. {Hudson}, R.~J. {Smith}, J.~R. {Lucey}, D.~J. {Schlegel}, and R.~L.
  {Davies}, {\it {A Large-scale Bulk Flow of Galaxy Clusters}},  {\em \apjl}
  {\bf 512} (Feb., 1999) L79--L82,
  [\href{http://arxiv.org/abs/astro-ph/9901001}{{\tt astro-ph/9901001}}].

\bibitem{Willick1999}
J.~A. {Willick}, {\it {The Las Campanas/Palomar 10,000 Kilometer Per Second
  Cluster Survey. I. Properties of the Tully-Fisher Relation}},  {\em \apj}
  {\bf 516} (May, 1999) 47--61,
  [\href{http://arxiv.org/abs/astro-ph/9809160}{{\tt astro-ph/9809160}}].

\bibitem{CouWilStr2000}
S.~{Courteau}, J.~A. {Willick}, M.~A. {Strauss}, D.~{Schlegel}, and
  M.~{Postman}, {\it {Shellflow. I. The Convergence of the Velocity Field at
  6000 Kilometers Per Second}},  {\em \apj} {\bf 544} (Dec., 2000) 636--640,
  [\href{http://arxiv.org/abs/astro-ph/0002420}{{\tt astro-ph/0002420}}].

\bibitem{WatFel2007}
R.~{Watkins} and H.~A. {Feldman}, {\it {Power spectrum shape from peculiar
  velocity data}},  {\em \mnras} {\bf 379} (July, 2007) 343--348,
  [\href{http://arxiv.org/abs/astro-ph/0702751}{{\tt astro-ph/0702751}}].

\bibitem{DavScr2014}
T.~M. {Davis} and M.~I. {Scrimgeour}, {\it {Deriving accurate peculiar
  velocities (even at high redshift)}},  {\em \mnras} {\bf 442} (Aug., 2014)
  1117--1122, [\href{http://arxiv.org/abs/1405.0105}{{\tt arXiv:1405.0105}}].

\bibitem{Nusser2014}
A.~{Nusser}, {\it {An Inconsistency in the Standard Maximum Likelihood
  Estimation of Bulk Flows}},  {\em \apj} {\bf 795} (Nov., 2014) 3,
  [\href{http://arxiv.org/abs/1405.6271}{{\tt arXiv:1405.6271}}].

\bibitem{Kashlinsky1991}
A.~{Kashlinsky} and B.~J.~T. {Jones}, {\it {Large-scale structure in the
  universe}},  {\em \nat} {\bf 349} (Feb., 1991) 753--760.

\bibitem{BosJiaHea2014}
F.~C.~v.~d. {Bosch}, F.~{Jiang}, A.~{Hearin}, D.~{Campbell}, D.~{Watson}, and
  N.~{Padmanabhan}, {\it {Coming of age in the dark sector: how dark matter
  haloes grow their gravitational potential wells}},  {\em \mnras} {\bf 445}
  (Dec., 2014) 1713--1730, [\href{http://arxiv.org/abs/1409.2750}{{\tt
  arXiv:1409.2750}}].

\bibitem{Komatsu2009}
E.~{Komatsu}, J.~{Dunkley}, M.~R. {Nolta}, C.~L. {Bennett}, B.~{Gold},
  G.~{Hinshaw}, N.~{Jarosik}, D.~{Larson}, M.~{Limon}, L.~{Page}, D.~N.
  {Spergel}, M.~{Halpern}, R.~S. {Hill}, A.~{Kogut}, S.~S. {Meyer}, G.~S.
  {Tucker}, J.~L. {Weiland}, E.~{Wollack}, and E.~L. {Wright}, {\it {Five-Year
  Wilkinson Microwave Anisotropy Probe Observations: Cosmological
  Interpretation}},  {\em \apjs} {\bf 180} (Feb., 2009) 330--376,
  [\href{http://arxiv.org/abs/0803.0547}{{\tt arXiv:0803.0547}}].

\bibitem{Pike2005}
R.~W. {Pike} and M.~J. {Hudson}, {\it {Cosmological Parameters from the
  Comparison of the 2MASS Gravity Field with Peculiar Velocity Surveys}},  {\em
  \apj} {\bf 635} (Dec., 2005) 11--21,
  [\href{http://arxiv.org/abs/astro-ph/0511012}{{\tt astro-ph/0511012}}].

\bibitem{Abate2009}
A.~{Abate} and P.~{Erdo{\v g}du}, {\it {Peculiar velocities into the next
  generation: cosmological parameters from the SFI++ survey}},  {\em \mnras}
  {\bf 400} (Dec., 2009) 1541--1547,
  [\href{http://arxiv.org/abs/0905.2967}{{\tt arXiv:0905.2967}}].

\bibitem{NusBraDav2011}
A.~{Nusser}, E.~{Branchini}, and M.~{Davis}, {\it {Bulk Flows from Galaxy
  Luminosities: Application to 2Mass Redshift Survey and Forecast for
  Next-generation Data Sets}},  {\em \apj} {\bf 735} (July, 2011) 77--+,
  [\href{http://arxiv.org/abs/1102.4189}{{\tt arXiv:1102.4189}}].

\bibitem{NusDav2011}
A.~{Nusser} and M.~{Davis}, {\it {The Cosmological Bulk Flow: Consistency with
  {$\Lambda$}CDM and z ${\approx}$ 0 Constraints on {$\sigma$}$_{8}$ and
  {$\gamma$}}},  {\em \apj} {\bf 736} (Aug., 2011) 93,
  [\href{http://arxiv.org/abs/1101.1650}{{\tt arXiv:1101.1650}}].

\bibitem{WatFelHud2009}
R.~{Watkins}, H.~A. {Feldman}, and M.~J. {Hudson}, {\it {Consistently large
  cosmic flows on scales of 100h$^{-1}$Mpc: a challenge for the standard
  {$\Lambda$}CDM cosmology}},  {\em \mnras} {\bf 392} (Jan., 2009) 743--756,
  [\href{http://arxiv.org/abs/0809.4041}{{\tt arXiv:0809.4041}}].

\bibitem{FelWatHud2010}
H.~A. {Feldman}, R.~{Watkins}, and M.~J. {Hudson}, {\it {Cosmic flows on 100
  h$^{-1}$ Mpc scales: standardized minimum variance bulk flow, shear and
  octupole moments}},  {\em \mnras} {\bf 407} (Oct., 2010) 2328--2338,
  [\href{http://arxiv.org/abs/0911.5516}{{\tt arXiv:0911.5516}}].

\bibitem{MaGorFel2011}
Y.-Z. {Ma}, C.~{Gordon}, and H.~A. {Feldman}, {\it {Peculiar velocity field:
  Constraining the tilt of the Universe}},  {\em \prd} {\bf 83} (May, 2011)
  103002--+, [\href{http://arxiv.org/abs/1010.4276}{{\tt arXiv:1010.4276}}].

\bibitem{Turnbull2012}
S.~J. {Turnbull}, M.~J. {Hudson}, H.~A. {Feldman}, M.~{Hicken}, R.~P.
  {Kirshner}, and R.~{Watkins}, {\it {Cosmic flows in the nearby universe from
  Type Ia supernovae}},  {\em \mnras} {\bf 420} (Feb., 2012) 447--454,
  [\href{http://arxiv.org/abs/1111.0631}{{\tt arXiv:1111.0631}}].

\bibitem{MacFelFer2012}
E.~{Macaulay}, H.~A. {Feldman}, P.~G. {Ferreira}, A.~H. {Jaffe}, S.~{Agarwal},
  M.~J. {Hudson}, and R.~{Watkins}, {\it {Power spectrum estimation from
  peculiar velocity catalogues}},  {\em \mnras} {\bf 425} (Sept., 2012)
  1709--1717, [\href{http://arxiv.org/abs/1111.3338}{{\tt arXiv:1111.3338}}].

\bibitem{WatFel2014}
R.~{Watkins} and H.~A. {Feldman}, {\it {An Unbiased Estimator of Peculiar
  Velocity with Gaussian Distributed Errors for Precision Cosmology}},  {\em
  ArXiv e-prints} (Nov., 2014) [\href{http://arxiv.org/abs/1411.6665}{{\tt
  arXiv:1411.6665}}].

\bibitem{WatFel2015}
R.~{Watkins} and H.~A. {Feldman}, {\it {Large-scale bulk flows from the
  Cosmicflows-2 catalogue}},  {\em \mnras} {\bf 447} (Feb., 2015) 132--139,
  [\href{http://arxiv.org/abs/1407.6940}{{\tt arXiv:1407.6940}}].

\bibitem{Kaiser1988}
N.~{Kaiser}, {\it {Theoretical implications of deviations from Hubble flow}},
  {\em \mnras} {\bf 231} (Mar., 1988) 149--167.

\bibitem{MacFelFer2011}
E.~{Macaulay}, H.~{Feldman}, P.~G. {Ferreira}, M.~J. {Hudson}, and
  R.~{Watkins}, {\it {A slight excess of large-scale power from moments of the
  peculiar velocity field}},  {\em \mnras} {\bf 414} (June, 2011) 621--626,
  [\href{http://arxiv.org/abs/1010.2651}{{\tt arXiv:1010.2651}}].

\bibitem{AgaFelWat2012}
S.~{Agarwal}, H.~A. {Feldman}, and R.~{Watkins}, {\it {Testing the minimum
  variance method for estimating large-scale velocity moments}},  {\em \mnras}
  {\bf 424} (Aug., 2012) 2667--2675,
  [\href{http://arxiv.org/abs/1201.0128}{{\tt arXiv:1201.0128}}].

\bibitem{HonSprStaScr2014}
T.~{Hong}, C.~M. {Springob}, L.~{Staveley-Smith}, M.~I. {Scrimgeour}, K.~L.
  {Masters}, L.~M. {Macri}, B.~S. {Koribalski}, D.~H. {Jones}, and T.~H.
  {Jarrett}, {\it {2MTF - IV. A bulk flow measurement of the local Universe}},
  {\em \mnras} {\bf 445} (Nov., 2014) 402--413,
  [\href{http://arxiv.org/abs/1409.0287}{{\tt arXiv:1409.0287}}].

\bibitem{ScrDavBlaSta2015}
M.~I. {Scrimgeour}, T.~M. {Davis}, C.~{Blake}, L.~{Staveley-Smith},
  C.~{Magoulas}, C.~M. {Springob}, F.~{Beutler}, M.~{Colless}, A.~{Johnson},
  D.~H. {Jones}, J.~{Koda}, J.~R. {Lucey}, Y.-Z. {Ma}, J.~{Mould}, and G.~B.
  {Poole}, {\it {The 6dF Galaxy Survey: bulk flows on 50-70 h$^{-1}$ Mpc
  scales}},  {\em \mnras} {\bf 455} (Jan., 2016) 386--401,
  [\href{http://arxiv.org/abs/1511.06930}{{\tt arXiv:1511.06930}}].

\bibitem{Li2012}
M.~{Li}, J.~{Pan}, L.~{Gao}, Y.~{Jing}, X.~{Yang}, X.~{Chi}, L.~{Feng},
  X.~{Kang}, W.~{Lin}, G.~{Shan}, L.~{Wang}, D.~{Zhao}, and P.~{Zhang}, {\it
  {Bulk Flow of Halos in {$\Lambda$}CDM Simulation}},  {\em \apj} {\bf 761}
  (Dec., 2012) 151, [\href{http://arxiv.org/abs/1207.5338}{{\tt
  arXiv:1207.5338}}].

\bibitem{RatKovItz2013}
B.~{Rathaus}, E.~D. {Kovetz}, and N.~{Itzhaki}, {\it {Studying the peculiar
  velocity bulk flow in a sparse survey of Type Ia SNe}},  {\em \mnras} {\bf
  431} (June, 2013) 3678--3684, [\href{http://arxiv.org/abs/1301.7710}{{\tt
  arXiv:1301.7710}}].

\bibitem{Dressler1987}
A.~{Dressler}, S.~M. {Faber}, D.~{Burstein}, R.~L. {Davies}, D.~{Lynden-Bell},
  R.~J. {Terlevich}, and G.~{Wegner}, {\it {Spectroscopy and photometry of
  elliptical galaxies - A large-scale streaming motion in the local universe}},
   {\em \apjl} {\bf 313} (Feb., 1987) L37--L42.

\bibitem{Lynden1989}
D.~{Lynden-Bell}, O.~{Lahav}, and D.~{Burstein}, {\it {Cosmological deductions
  from the alignment of local gravity and motion}},  {\em \mnras} {\bf 241}
  (Nov., 1989) 325--345.

\bibitem{Turner1992}
M.~S. {Turner}, {\it {The tilted universe}},  {\em General Relativity and
  Gravitation} {\bf 24} (Jan., 1992) 1--7.

\bibitem{KasAtrKoc2008}
A.~{Kashlinsky}, F.~{Atrio-Barandela}, D.~{Kocevski}, and H.~{Ebeling}, {\it {A
  Measurement of Large-Scale Peculiar Velocities of Clusters of Galaxies:
  Results and Cosmological Implications}},  {\em \apjl} {\bf 686} (Oct., 2008)
  L49--L52, [\href{http://arxiv.org/abs/0809.3734}{{\tt arXiv:0809.3734}}].

\bibitem{Larjo2010}
K.~{Larjo} and T.~S. {Levi}, {\it {Bubble, bubble, flow and Hubble: large scale
  galaxy flow from cosmological bubble collisions}},  {\em \jcap} {\bf 8}
  (Aug., 2010) 34, [\href{http://arxiv.org/abs/0910.4159}{{\tt
  arXiv:0910.4159}}].

\bibitem{Fialkov2010}
A.~{Fialkov}, N.~{Itzhaki}, and E.~D. {Kovetz}, {\it {Cosmological imprints of
  pre-inflationary particles}},  {\em \jcap} {\bf 2} (Feb., 2010) 4,
  [\href{http://arxiv.org/abs/0911.2100}{{\tt arXiv:0911.2100}}].

\bibitem{TulShaKar2008}
R.~B. {Tully}, E.~J. {Shaya}, I.~D. {Karachentsev}, H.~M. {Courtois}, D.~D.
  {Kocevski}, L.~{Rizzi}, and A.~{Peel}, {\it {Our Peculiar Motion Away from
  the Local Void}},  {\em \apj} {\bf 676} (Mar., 2008) 184--205,
  [\href{http://arxiv.org/abs/0705.4139}{{\tt arXiv:0705.4139}}].

\bibitem{Bland2014}
J.~{Bland-Hawthorn} and S.~{Sharma}, {\it {Void asymmetries in the cosmic web:
  a mechanism for bulk flows}},  {\em ArXiv e-prints} (Oct., 2014)
  [\href{http://arxiv.org/abs/1410.3845}{{\tt arXiv:1410.3845}}].

\bibitem{LasDamas}
C.~{McBride}, A.~{Berlind}, R.~{Scoccimarro}, R.~{Wechsler}, M.~{Busha},
  J.~{Gardner}, and F.~{van den Bosch}, {\it {LasDamas Mock Galaxy Catalogs for
  SDSS}},  {\em BAAS} {\bf 41} (Jan., 2009) 425.06.

\bibitem{HR}
J.~{Kim}, C.~{Park}, J.~R. {Gott}, III, and J.~{Dubinski}, {\it {The Horizon
  Run N-Body Simulation: Baryon Acoustic Oscillations and Topology of
  Large-scale Structure of the Universe}},  {\em \apj} {\bf 701} (Aug., 2009)
  1547--1559, [\href{http://arxiv.org/abs/0812.1392}{{\tt arXiv:0812.1392}}].

\bibitem{ZA70}
Y.~B. {Zel'dovich}, {\it {Gravitational instability: An approximate theory for
  large density perturbations.}},  {\em \aaps} {\bf 5} (Mar., 1970) 84--89.

\bibitem{ZA}
D.~J. {Eisenstein}, H.-J. {Seo}, E.~{Sirko}, and D.~N. {Spergel}, {\it
  {Improving Cosmological Distance Measurements by Reconstruction of the Baryon
  Acoustic Peak}},  {\em \apj} {\bf 664} (Aug., 2007) 675--679,
  [\href{http://arxiv.org/abs/astro-ph/0604362}{{\tt astro-ph/0604362}}].

\bibitem{DavEfsFre1985}
M.~{Davis}, G.~{Efstathiou}, C.~S. {Frenk}, and S.~D.~M. {White}, {\it {The
  evolution of large-scale structure in a universe dominated by cold dark
  matter}},  {\em \apj} {\bf 292} (May, 1985) 371--394.

\bibitem{GarConMcB2007}
J.~P. {Gardner}, A.~{Connolly}, and C.~{McBride}, {\it {A Framework for
  Analyzing Massive Astrophysical Datasets on a Distributed Grid}},  {\em
  \pasp} {\bf 376} (Oct., 2007) 69.

\bibitem{CroPueSco2006}
M.~{Crocce}, S.~{Pueblas}, and R.~{Scoccimarro}, {\it {Transients from initial
  conditions in cosmological simulations}},  {\em \mnras} {\bf 373} (Nov.,
  2006) 369--381, [\href{http://arxiv.org/abs/astro-ph/0606505}{{\tt
  astro-ph/0606505}}].

\bibitem{LSS}
P.~J.~E. {Peebles}, {\em {The large-scale structure of the universe}}.
\newblock Princeton, N.J., Princeton University Press, 1980, 1980.

\bibitem{Linder2005}
E.~V. {Linder}, {\it {Cosmic growth history and expansion history}},  {\em
  \prd} {\bf 72} (Aug., 2005) 043529,
  [\href{http://arxiv.org/abs/astro-ph/0507263}{{\tt astro-ph/0507263}}].

\end{thebibliography}\endgroup

\end{document}